\documentclass{wscpaperproc}
\usepackage{wscsetup}
\usepackage{fancyheadings,latexsym,url,caption2}
\usepackage{graphicx}
\usepackage{times,mathptmx}

\usepackage{amsmath,amsfonts,amssymb,amsbsy,amsthm}

\begin{document}

\pagestyle{fancyplain} \thispagestyle{plain}
\lhead{\fancyplain{\textit{Reducing the Variance of Likelihood Ratio Greeks in Monte Carlo }}{}}

 \chead{\fancyplain{}{\itshape Luca Capriotti}}

\headsep=9pt \rhead{} \cfoot{}\headrulewidth=0pt

\makeatletter
\let\@internalcite\cite
\def\cite{\def\@citeseppen{-1000}%
    \def\@cite##1##2{(##1\if@tempswa , ##2\fi)}%
    \def\citeauthoryear##1##2##3{##1 ##3}\@internalcite}
\def\citeNP{\def\@citeseppen{-1000}%
    \def\@cite##1##2{##1\if@tempswa , ##2\fi}%
    \def\citeauthoryear##1##2##3{##1 ##3}\@internalcite}
\def\citeN{\def\@citeseppen{-1000}%
    \def\@cite##1##2{##1\if@tempswa, ##2)\else{}\fi}%
    \def\citeauthoryear##1##2##3{##1 (##3)}\@citedata}
\def\citeA{\def\@citeseppen{-1000}%
    \def\@cite##1##2{(##1\if@tempswa , ##2\fi)}%
    \def\citeauthoryear##1##2##3{##1}\@internalcite}
\def\citeANP{\def\@citeseppen{-1000}%
    \def\@cite##1##2{##1\if@tempswa , ##2\fi}%
    \def\citeauthoryear##1##2##3{##1}\@internalcite}
\def\shortcite{\def\@citeseppen{-1000}%
    \def\@cite##1##2{(##1\if@tempswa , ##2\fi)}%
    \def\citeauthoryear##1##2##3{##2 ##3}\@internalcite}
\def\shortciteNP{\def\@citeseppen{-1000}%
    \def\@cite##1##2{##1\if@tempswa , ##2\fi}%
    \def\citeauthoryear##1##2##3{##2 ##3}\@internalcite}
\def\shortciteN{\def\@citeseppen{-1000}%
    \def\@cite##1##2{##1\if@tempswa, ##2\else{}\fi}%
    \def\citeauthoryear##1##2##3{##2 (##3)}\@citedata}
\def\shortciteA{\def\@citeseppen{-1000}%
    \def\@cite##1##2{(##1\if@tempswa , ##2\fi)}%
    \def\citeauthoryear##1##2##3{##2}\@internalcite}
\def\shortciteANP{\def\@citeseppen{-1000}%
    \def\@cite##1##2{##1\if@tempswa , ##2\fi}%
    \def\citeauthoryear##1##2##3{##2}\@internalcite}
\def\citeyear{\def\@citeseppen{-1000}%
    \def\@cite##1##2{(##1\if@tempswa , ##2\fi)}%
    \def\citeauthoryear##1##2##3{##3}\@citedata}
\def\citeyearNP{\def\@citeseppen{-1000}%
    \def\@cite##1##2{##1\if@tempswa , ##2\fi}%
    \def\citeauthoryear##1##2##3{##3}\@citedata}
%
%
%
\def\@citedata{%
    \@ifnextchar [{\@tempswatrue\@citedatax}%
                  {\@tempswafalse\@citedatax[]}%
}

\def\@citedatax[#1]#2{%
\if@filesw\immediate\write\@auxout{\string\citation{#2}}\fi%
  \def\@citea{}\@cite{\@for\@citeb:=#2\do%
    {\@citea\def\@citea{, }\@ifundefined
       {b@\@citeb}{{\bf ?}%
       \@warning{Citation `\@citeb' on page \thepage \space undefined}}%
{\csname b@\@citeb\endcsname}}}{#1}}%

%
\def\@citex[#1]#2{%
\if@filesw\immediate\write\@auxout{\string\citation{#2}}\fi%
  \def\@citea{}\@cite{\@for\@citeb:=#2\do%
    {\@citea\def\@citea{, }\@ifundefined
       {b@\@citeb}{{\bf ?}%
       \@warning{Citation `\@citeb' on page \thepage \space undefined}}%
{\csname b@\@citeb\endcsname}}}{#1}}%

%
\def\@biblabel#1{}
\makeatother


\renewcommand{\refname}{REFERENCES}

\newdimen\bibindent
\bibindent=0.0em
\def\thebibliography#1{\section*{\refname}\list
   {}{\settowidth\labelwidth{[#1]}
   \leftmargin\parindent
   \itemindent -\parindent
   \listparindent \itemindent
   \itemsep 0pt
   \parsep 0pt}
   \def\newblock{}
   \sloppy
   \sfcode`\.=1000\relax}


\title{\bf \vspace*{-0.02in} REDUCING THE VARIANCE OF LIKELIHOOD RATIO GREEKS IN MONTE CARLO}

\author{Luca\ Capriotti                       \\ [12pt]
        Global Modelling and Analytics Group \\
        Credit Suisse Group, Investment Banking Division \\
        Eleven Madison Avenue, New York, NY 10010, U.S.A.
}

\maketitle

\section*{ABSTRACT}

We investigate the use of Antithetic Variables, Control Variates and Importance Sampling
to reduce the statistical errors of option sensitivities calculated
with the Likelihood Ratio Method in Monte Carlo. We show how Antithetic Variables
solve the well-known problem of the divergence of the variance of Delta
for short maturities and small volatilities. With numerical examples within a Gaussian Copula framework,
we show how simple Control Variates and Importance Sampling strategies provide computational
savings up to several orders of magnitude.

\section{INTRODUCTION}  \label{s:intro}

Monte Carlo (MC) simulations are one the main tools employed in
the Financial Services industry for pricing and hedging derivatives
securities.  In fact, as a result of the ever increasing level of sophistication of
the financial markets, a considerable fraction of the pricing
models employed by investment firms is too complex to be
treated by analytic or deterministic numerical methods. For these
models, MC simulation is the only computationally feasible pricing
method.

The main drawback of MC methods is that they are generally
computationally expensive. These efficiency issues become
even more dramatic when MC simulations are used for the calculation of
price sensitivities, i.e., the derivatives of the option price
with respect to the parameters of the underlying model, also known
as {\em Greeks}.  In fact, the
standard method for the calculation of a price sensitivity, say
with respect to a parameter $\theta_k$, is based on a finite difference
approximation of the derivatives $\partial V(\theta) /\partial
\theta_k$. This method, also known as `bump and reval', involves
repeating the MC simulation,  and evaluating
the finite difference estimate
\begin{equation}\label{bumping}
\frac{\partial V(\theta)}{\partial \theta_k} \simeq
\frac{V(\theta_1,\ldots, \theta_k + \Delta \theta,\ldots,
\theta_n) - V(\theta)}{\Delta \theta}~,
\end{equation}
for a small increment $\Delta \theta$. The main virtue of this
method is that it is straightforward to understand, and it
requires minimal implementation effort. The drawback is that
additional MC simulations are required for each sensitivity, and
that the finite differences (\ref{bumping}) may be affected by large
statistical errors, especially for payout with discontinuities \cite{GlassMCbook}.
As a result, hedging derivative securities with MC simulations
can be extremely time consuming.

Alternative methods for the calculation of price sensitivities have been proposed
in the literature (for a review see \citeNP{GlassMCbook}). Here we concentrate
on the so-called Likelihood Ratio Method (LRM).  The principal advantage of this technnique
when compared to `bump and reval' is that it allows to calculate
all the sensitivities simultaneously in a single MC simulation, and a single set of
payout evaluations. In addition, the variance properties of LRM estimators are not
affected as much by discontinuities in the payoff.
As a result, for digital and barrier options, LRM may provide a better convergence 
than bumping \cite{GlassMCbook}. The main drawback is that the statistical 
uncertainties of LRM estimators are nonetheless difficult to predict, and can be 
sometimes large.
What is worse, in some cases such uncertainties are even
known to diverge thus making the MC simulation very time consuming if not
hopeless in practice.

In order to address this difficulty, in this paper we investigate three
Variance Reduction techniques -- Antithetic Variables, Control Variates,
and Importance Sampling -- that can dramatically improve the MC convergence of LRM estimators.
In the next Section, we begin by reviewing the rationale of LRM, specializing our discussion to a
Gaussian Copula framework very common in the financial practice. The use of
Antithetic Variables is discussed in Section \ref{AV}.  In particular, we will show
how this simple technique solves the well-known problem of the divergence of the
variance of LRM Deltas for short maturity and low volatility.
Then, in Section \ref{CVIS} we illustrate how Control Variates and
Importance Sampling can drastically suppress the statistical uncertainties of the
LRM estimators thus reducing the computational cost for the Greeks by orders
of magnitude.

\section{LIKELIHOOD RATIO METHOD}

The arbitrage-free price of a derivative security can be 
expressed in general as the expectation value of the discounted cash flows,
$G(x)$, over a risk-neutral probability density \cite{HarrKreps}, $P_\theta(x)$,
\begin{equation}\label{price}
V(\theta) = \mathbb{E}_P[G(x)] = \int\,dx \,\,G(x)\, P_\theta(x)~,
\end{equation}
where $x=(x_1,\ldots,x_N)$ is a $N$-dimensional vector representing the underlying random
factors upon which the claim is contingent. Here the vector $\theta=(\theta_1,\dots,\theta_n)$
represents a set of parameters whose value is generally determined by calibrating
the chosen model or, equivalently, the density $P_\theta(x)$,
on the prices of securities liquidly traded in the market.

Whenever the dimension $N$ of the state variable $x$ is large (say
$N \gtrsim 4$) MC methods are the only feasible
route for estimating expectation values of the form
(\ref{price}). In their simplest incarnation, these consist in
averaging the payout function $G(x)$ over $N_{p}$ independent
random realizations of the vector $x$, say $x[m]$, generated according to the
probability density $P_\theta(x)$,
\begin{equation}\label{mcprice}
V(\theta) \simeq \bar V = \frac{1}{N_{p}} \sum_{m=1}^{N_p} G(x[m])~.
\end{equation}
In fact, the central limit theorem \cite{CLT} ensures that,
for big enough samples, the values of the estimator $\bar V$ are
normally distributed around the true value, and converge for $N_p
\to \infty$ towards $V$ namely
\begin{equation}\label{sqrt}
V \simeq \frac{1}{N_p} \sum_{m=1}^{N_p} G(x[m]) \pm
\frac{\kappa}{\sqrt{N_p}}~,
\end{equation}
where $\kappa^2 =
E_P\left[G(x)^2\right]-E_P\left[G(x)\right]^2$ is
the variance of the MC estimator.
Here, following a common terminology, we refer to $\kappa/\sqrt{N_p}$ as {\em statistical uncertainty} or {\em statistical error}.
Although Eq.~(\ref{sqrt}) ensures the convergence of the MC average to the expectation value
(\ref{price}) provided that $\kappa$ is finite,
the square root law in (\ref{sqrt}) can make the calculation of accurate
estimates time consuming. 

This is particularly true, for the MC calculation of the Greeks.  In fact, the variance and bias properties
of finite difference estimators of the form (\ref{bumping}) can be in some cases rather poor \cite{GlassMCbook}.
This is because, while the bias of the finite difference (\ref{bumping}) can be made in general
arbitrarily small by reducing the value of $\Delta \theta$, its statistical error
can in some common cases diverge for $\Delta \theta \to 0$. When this happens, choosing a
value of $\Delta \theta$ small enough to reduce the bias to an acceptable level may require a large
computational cost in order to obtain statistically accurate results. For a convergence analysis see
e.g., \cite{milstein05}.

Several methods have been recently proposed in the literature in order to speed up the
calculation of option sensitivities \cite{GlassMCbook}. Here we will concentrate on the so-called
Likelihood Ratio Method (LRM).  Under mild regularity conditions on the probability
density $P_\theta(x)$, the sensitivity of the option price (\ref{price}) with respect to any parameter
$\theta_k$ can be obtained as
\begin{equation}\label{lrmeq}
\bar \theta_k = \frac{\partial V(\theta)}{\partial \theta_k} = \mathbb{E}_P[G(x)\Omega_k(x)]~,
\end{equation}
i.e., by calculating the expectation value of the original payout function multiplied by the so-called
{\em Likelihood Ratio} weight
\begin{equation}\label{LRMweight}
\Omega_k(x) = \frac{\partial \log P_\theta(x)}{\partial \theta_k}~,
\end{equation}
giving as MC estimator:
\begin{equation}\label{mcdelta}
\bar \theta_k \simeq \frac{1}{N_{p}} \sum_{m=1}^{N_p} G(x[m]) \Omega_k(x[m]) ~.
\end{equation}

\begin{figure}
\includegraphics[width=0.43\textwidth]{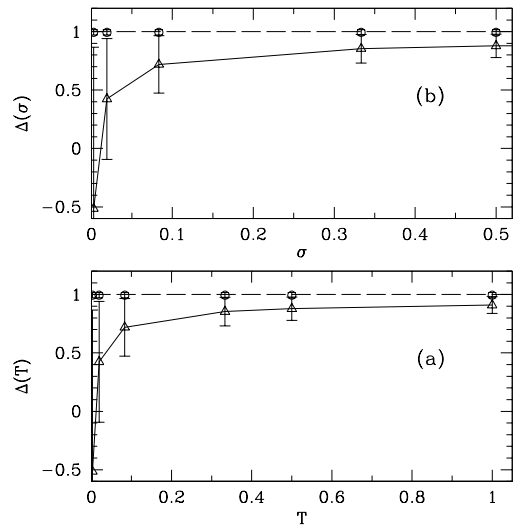}
\vspace{-5mm} \caption{\label{fig1} Delta of a European Call option (\ref{call}) for  $S_0 = 100$, $K=25$, and $r=0.05$
as a function of the time to maturity $T$ for $\sigma = 0.1$ (a),
and as a function of the volatility $\sigma$ for $T=1/12$ (b): crude MC (triangles and continuous line),
MC with Antithetic Variables (circles and dashed line).
}
\end{figure}

In the following, we will specialize our discussion to the case
where the probability distribution $P_\theta(x)$ is a $N$-dimensional
Gaussian Copula.  The latter is defined by a correlation matrix $\Sigma$, and a
set of $N$ marginal cumulative distributions $M_i(x_i)$, so that the joint
distribution reads
\begin{eqnarray}\label{gaucop}
F(x) &=& \prod_{i=1}^N \int_{-\infty}^{x_i} dy_i
P(y_1,\ldots,y_N) \nonumber \\ &=&
\Phi_N\left(\Phi^{-1}\left(M_1(x_1)\right),.,\Phi^{-1}\left(M_N(x_N)\right);\Sigma\right).
\end{eqnarray}
Here $\Phi_N(x_1,\ldots,x_N;\Sigma)$ is the cumulative
distribution of a $N$-dimensional Gaussian random  variable with
zero mean and correlation matrix $\Sigma$, and $\Phi(.)$ is the
standard normal cumulative distribution.

If we restrict to the case in which the correlation matrix
$\Sigma$ does {\em not} depend on the parameters $\theta$, the LRM weight
(\ref{LRMweight}) for the Gaussian Copula distribution (\ref{gaucop})
can be expressed as
\begin{equation}\label{weight}
\Omega_\theta(x) = \sum_{i=1}^{N} \partial_\theta \log m_i(x_i) -
Z(x)^T \, (\Sigma^{-1}-I) \,\partial_\theta Z(x)~,
\end{equation}
where $I$ is the $N$-dimensional identity matrix,
$m_i(x_i) = d M(x_i)/d x_i$ is the $i$-th marginal density
function, and the components of the vector $Z(x)$ are
\begin{equation}\label{zeta}
Z_i = \Phi^{-1}(M_i(x_i)),
\end{equation}
and those of $\partial_\theta Z(x)$ read
\begin{equation}\label{dcdf}
\partial_\theta Z_i = \frac{\partial_\theta
M_i(x_i)}{\phi(\Phi^{-1}(M_i(x_i)))}~,
\end{equation}
with $\phi(.)$ the standard normal density function.
A derivation of Eq.~(\ref{weight}) is given in the Appendix.

Gaussian Copula models of the form (\ref{gaucop}) are widely used in Financial Engineering. Indeed, the above formulation can
be used to evaluate structured European options written on several assets, e.g., equity,
commodity, rates or foreign exchange pairs \cite{hull}. In this case, the marginal distribution of each asset
is typically implied from liquidly traded Vanilla options, and the co-dependence
between the factors is modeled by means of the Gaussian Copula. Structured credit pricing, e.g.,
for CDO and CDO$^2$  \cite{schonbucher}, can be also performed within a similar framework.
In general, whenever the marginal distributions above are not known in closed form, e.g.,
they are calculated numerically by means of a calibration procedure, the derivatives in Eq.~(\ref{dcdf})
can be easily computed by means of finite differences. This does not
generally introduce accuracy or stability problems provided the calibration algorithms employed are
numerically stable.

It is easy to see that the LRM weights above give the expected result in the case a
multi-asset lognormal model of the form
\begin{equation}\label{lnmodel}
S_i = S_i^0\, \exp{\big[(r-\sigma_i^2/2)T+\sigma_i\sqrt{T}\,Z_i}\big]~,
\end{equation}
where $S_i^0$ and $\sigma_i$ are the spot price and volatility of the $i$-th asset,
$T$ is the maturity of the option, and $Z_i$ are standard
normal variables with correlation $\Sigma_{ij} = \mathbb{E}[Z_iZ_j]$. In fact, one clearly has $x_i = Z_i$ with
\begin{equation}
Z_i = \frac{\log{S_i}/{S_i^0} - (r-\sigma_i^2/2)T}{\sigma_i \sqrt{T}}
\end{equation}
so that the LRM weights for the $i$-th Delta and Vega \cite{hull} read respectively
\begin{equation}\label{deltalognormal}
\Omega_{\Delta}^i(Z) = \frac{\partial}{\partial S_i^0} \log P_\theta(x) = \frac{[\Sigma^{-1}Z]_i}{\sigma_i \sqrt{T} S_i^0} ~,
\end{equation}
and
\begin{equation}\label{vegalognormal}
\Omega_{V}^i(Z) = \frac{\partial}{\partial \sigma_i} \log P_\theta(x) = \Big(\frac{Z_i}{\sigma_i}-\sqrt{T}\Big)[\Sigma^{-1}Z]_i -\frac{1}{\sigma_i}~.
\end{equation}
It is straightforward to verify that these equations are in agreement with the general expression
for the LRM weight in a Gaussian model given in \cite{GlassMCbook}
\begin{eqnarray}\label{general}
\Omega_\theta(Z)&=& -\frac{1}{2} {\rm Tr}
\left[\hat\Sigma^{-1} \partial_{\theta}{\hat\Sigma}\,\right] +
\frac{1}{2}X \, \hat\Sigma^{-1} \,(\partial_{\theta} {\hat\Sigma}) \,
\hat\Sigma^{-1}X \nonumber \\ &+& X\,\hat\Sigma^{-1} \,
\partial_{\theta} \,{m}~,
\end{eqnarray}
with $m_i = \log{S_i^0} + (r-\sigma^2/2)T$, $X_i=\sigma_i\sqrt{T}\,Z_i$,
and $\hat\Sigma_{ij}=\sigma_i\sigma_j\Sigma_{ij}$.

In the special case of a single asset, the weights above simplify to the well-known expressions
\begin{equation}\label{deltasingle}
\Omega_{\Delta}(Z) = \frac{Z}{\sigma \sqrt{T} S^0}~,
\end{equation}
and
\begin{equation}\label{vegasingle}
\Omega_{V}(Z)= \frac{Z^2 - 1}{\sigma} - Z\sqrt{T}~.
\end{equation}

\section{SOLVING THE PROBLEM OF DELTA'S DIVERGING VARIANCE WITH ANTITHETIC VARIABLES}
\label{AV}

For a given number of MC iterations the calculation of
the Greeks by means of LRM is generally fast when compared to
bumping. However, the speed of convergence of the LRM estimators
is difficult to predict {\em a priori} for a given problem, it is payout and parameters
dependent, and can be in some instances particularly slow \cite{GlassMCbook}.
In fact, since the LRM weight has in general zero mean as a result of the identity
\begin{equation}
\partial_{\theta} \int dx \,\, P_\theta(x) = 0~,
\end{equation}
the LRM estimators for the Greeks have no definite sign. This can give rise to
poor variance properties whenever the configurations with opposite sign have similar weight
in the MC average (\ref{mcdelta}) so that the final outcome is the result of the
cancellation of two comparable and not necessarily highly correlated quantities.

\begin{table}[htb]
\caption{Delta of the Basket Call option (\ref{basketcall}) for $T=0.5$ and $r=0.05$.
The volatilities of the assets are all equal to $\sigma=0.3$,
and their Forwards range between 51.3 and 55.9. The uncertainties are reported in parenthesis. \label{tab:deltabasket}}
\begin{center}
\begin{tabular}{ccccc}
\hline
K   & AV        & AV+CV        & AV+LSIS \\ \hline
30  & 7.5(5)    & 7(1)$10^{4}$   & 1200(100) \\
40  & 3.5(3)    & 1200(100)    & 200(10)   \\
50  & 2.2(2)    & 410(3)       & 100(10)   \\
60  & 2.1(1)    & 11(1)        & 120(10)   \\
70  & 2.2(2)    & 3.1(3)       & 200(20)   \\
80  & 1.9(2)    & 2.1(3)       & 610(70)   \\
\hline
\end{tabular}
\end{center}
\end{table}

In particular, a common problem generally reported in the literature \cite{GlassMCbook,Jaeckelbook}
is the divergence of the variance of the LRM weight for Delta Eqs. (\ref{deltalognormal}) and (\ref{deltasingle})
in the limit of small volatility and short maturity. Indeed, from Eq.~(\ref{deltalognormal})
\begin{equation}
{\rm Var}[\Omega_{\Delta}^i] \propto \frac{1}{\sigma_i^2 T} \to \infty
\end{equation}
for $\sigma_i \sqrt{T}\to 0$.  This is illustrated in Figure 1 for a 
`deep in the money' Call option (i.e., with very low strike compared to the expectation value of the underlying asset, or Forward, see \citeNP{hull}) on a single lognormal asset with undiscounted payout
\begin{equation}\label{call} 
G(S) = (S-K)^+
\end{equation}
where $K$ is the strike price. The LRM estimator becomes extremely noisy
for $T\to 0$ and for small volatility, to the point of not providing any useful information for
maturities shorter than a few weeks for any practical number of MC iterations.

The divergence of the LRM weight for Delta is due to the break down of the absolute continuity property
of the probability density function, which is required to take the derivative inside the expectation
in Eq.~(\ref{lrmeq}) \cite{GlassMCbook}. This generally affects the LRM estimators
(\ref{weight}) for Delta also for non-lognormal models.

Although, to the best of our knowledge, it was not previously noted in the literature,
this problem can be easily overcome by using Antithetic Variables.
Indeed, since the LRM weight (\ref{deltalognormal}) is {\em odd} in each of the
Gaussian random variable $Z_i$, the Antithetic estimator  \cite{GlassMCbook} for the weight reads
\begin{equation}\label{anthest}
\Omega_{\Delta}^i |_{ant} = \frac{\Omega_\Delta^i(Z)+\Omega_\Delta^i(-Z)}{2} \equiv 0
\end{equation}
so that $\mathbb{E}_P[\Omega_{\Delta}^i|_{ant}] = 0$, with {\em zero variance}. What is
more, it is also possible show that the variance of the product
of the payout and the weight in (\ref{mcdelta}) is generally bounded as $\sigma \sqrt{T}\to 0$. This can be
realized by means of the following simple heuristic argument. Consider for simplicity a
single asset payout under the lognormal model (\ref{lnmodel}).
This can be approximated for small maturities and volatilities as
\begin{equation}
P(Z) \simeq c_0 + c_1 \sigma \sqrt{T}\,Z + O(\sigma^2 T)
\end{equation}
with $c_0$ and $c_1$ constants . As a result, the LRM Antithetic estimator for Delta (\ref{mcdelta}) reads in this
limit
\begin{equation}
\Omega_\Delta|_{ant} \simeq c_0\frac{Z-Z}{2\sigma \sqrt T S^0} + c_1 \frac{Z^2}{S^0} + O(\sigma \sqrt{T})
\end{equation}
whose variance is clearly bounded for $\sigma \sqrt{T} \to 0$ as a result of the cancellation of the
leading term.

\begin{figure}
\vspace{-25mm}
\includegraphics[width=0.43\textwidth]{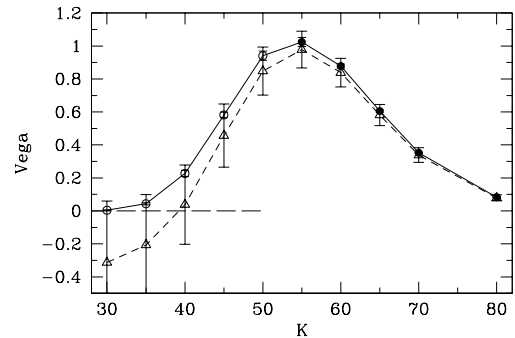}
\vspace{-5mm} \caption{\label{fig2} Vega of the Basket Call option (\ref{basketcall})
as a function of the strike price. Crude MC: triangles and dashed line. MC with Variance Reduction techniques:
circles and continuous line. Empty circles: Antithetic Variables with Control Variates. Full Circles: Antithetic Variables with LSIS.
}
\end{figure}

Figure \ref{fig1} illustrates the efficacy of the method: the Antithetic LRM
estimator provides a stable Delta and a practically constant statistical uncertainty
for all maturities and volatilities. The variance reduction with respect to the
crude LRM estimators is around 10 for 1 year, and around 500 for 1
week maturity. 

While the results presented here are for a lognormal model, we have found comparable variance reductions also for market-implied marginal 
distributions $M_i(x)$ for a variety of equity, foreign exchange, interest rate and commodity underlying assets. Indeed, 
the antisymmetry of the LRM weight (\ref{weight}) for Delta is generally satisfied, at least approximately, also for skewed 
distributions.  It is also worth noting that although here we have limited our discussions to European-style payouts in a Gaussian Copula
framework, it is easy to realize that Antithetic Variables generally solve the problem of the divergence of the
variance of Delta estimators also for path-dependent options 
(similar ideas have been also used by Mike Giles in the implementation of the so-called Vibrato Monte Carlo technique \cite{GilesVibrato}). 

\section{CONTROL VARIATES AND IMPORTANCE SAMPLING}
\label{CVIS}

Although the variance of LRM estimators generally remains finite when using Antithetic Variables,
their convergence can still be poor in some cases, especially for Vega. For this reason,
an efficient implementation of LRM generally requires the use of other Variance Reduction
strategies. Here we consider Control Variates and Importance Sampling \cite{GlassMCbook}.

The Control Variates method aims to reduce the statistical uncertainty of a MC average by
exploiting the correlation of its statistical samples with those of some quantity (the control) whose expectation
value is known {\em a priori} \cite{GlassMCbook}. This technique may result in spectacular variance
reductions but requires some closely related estimator with a known integral over
the sampled probability distribution.

For this discussion, we will restrict ourselves to
controls that in our context are generally readily available, and we will consider the
LRM weights and the derivatives of the first moment of the marginal distributions in
Eq.~(\ref{gaucop}). In fact, as previously mentioned, the LRM weight (\ref{LRMweight})
has always zero expectation value. It is also reasonable to expect the weight to be
somewhat correlated with the corresponding Greek estimator in (\ref{mcdelta}). On the other hand, the first moment of
each marginal distribution is usually known as it contains the information on the Forward of
the underlying asset, which is the first thing to be usually matched with the available market data.
Its derivatives with respect to the main model parameters are also generally known.
For instance, in a typical financial context, including models with skew \cite{hull}, one has
\begin{equation}\label{deltacv}
\frac {\partial  \mathbb{E}  [{S_i}] }{\partial S_i^0} = P(0,T)^{-1}~,
\end{equation}
where $P(0,T)$ is today's ($t = 0$) price  of a zero coupon bond maturing at time $T$, and
\begin{equation}\label{vegacv}
\frac {\partial  \mathbb{E}  [{S_i}] }{\partial \sigma_i} = 0~.
\end{equation}
In the following, when calculating the Greeks of the $i$-th asset, we will use the
corresponding LRM weight, and Risk  of the Forward. Note, however, that when calculating the LRM
Delta with Antithetic Variables, we can only use the Delta of the Forward as a control, as the
Antithetic estimator for the LRM weight (\ref{anthest}) is identically zero.

\begin{table}[h]
\caption{Same of Table \ref{tab:deltabasket} for Vega.\label{tab:vegabasket}}
\begin{center}
\begin{tabular}{cccc}
\hline
K   &  AV       &  AV+CV      & AV+LSIS \\ \hline
30  &  1.3(1)   &  4(1)$10^4$ & 11(1)   \\
40  &  1.7(1)   &  150(10)    & 7.0(6)   \\
50  &  2.1(1)   &   22(2)     & 8.0(8)   \\
60  &  2.1(2)   &  5.5(6)     & 80(10)    \\
70  &  1.9(2)   &  5.2(5)     & 340(40)   \\
80  &  1.7(3)   &  1.7(3)     & 1100(100) \\
\hline
\end{tabular}
\end{center}
\end{table}

Importance Sampling techniques, on the other hand, do not rely on the knowledge of any closely
correlated estimator but aim to reduce the variance by sampling more effectively the domain of
integration in Eq.~(\ref{price}). Here we will use a recently introduced Importance Sampling
strategy based on a Least-squares optimization, namely the Least-Squares Importance Sampling (LSIS)
\cite{lsis1,lsis2}. In particular we will use as trial densities mean-shifted single mode and bi-mode
multivariate Gaussian distributions (see references above).

In Table \ref{tab:deltabasket}, we compare the results obtained by using Antithetic Variables only,
 and in combination with Control Variates and Importance Sampling for the Delta of one of
 $N=10$ assets of a Basket Call option with undiscounted payout
\begin{equation}\label{basketcall}
G(S) = \Big(\frac{1}{N} \sum_{i=1}^N S_i - K\Big)^+~,
\end{equation}
in a lognormal model of the form (\ref{lnmodel}). Here, as an indicator of the
efficiency gains introduced by the different methods, we
have defined the variance (efficiency) ratio as
\begin{equation}
{\rm ER} = \left(\frac{\sigma[{\rm Crude MC}]}{\sigma[{\rm VR}]}\right)^2
\end{equation}
where the numerator and denominator are respectively the statistical errors (for the same number
of MC paths) of the crude estimator and of the one obtained with different Variance Reduction techniques.

As shown in Table \ref{tab:deltabasket}, for the considered maturity, the efficacy of Antithetic Variables decreases moving away from
the deep in the money region,
and is generally limited to around a factor of two for larger strikes. Using the Delta of
the corresponding Forward (\ref{deltacv}) as Control Variate provides about an order
of magnitude variance reduction around the `at the money' point (i.e., for strikes around the Forward value).
Decreasing further the strike results in a spectacular suppression of the statistical errors of the Control Variate estimator.
This is expected due to the very high correlation between the option payoff in this regime and a simple
Forward contract. However, as the strike moves instead in the `out of the money' region
(high strikes compared to the Forward) such correlation rapidly decreases, and Control Variates become practically ineffective.

On the other hand, a simple Importance Sampling strategy based on a single mean-shifted Gaussian
trial density and implemented by means of the LSIS approach proves to be
very effective at all level of moneyness. In fact,  although Importance Sampling gives
smaller efficiency gains than Control Variates for deep in the money options,
it provides at least two orders of magnitude speed up across all strikes,
including the out of the money regions where Control Variates lose their efficacy.
Indeed, as it can be generally expected when using mean-shifted trial densities \cite{lsis1}, Importance Sampling
is particularly effective moving away from the money in either direction.

LRM estimators for Vega  are generally noisier than the ones for Delta. This can be understood
from the form of the weight in Eq.~(\ref{vegalognormal}) as it involves the second moments of the Random
increments which have generally a larger variance. This is illustrated in Fig.~\ref{fig2} where we
plot one of the Vegas for the Basket option above as a function of the strike: for low strikes, as Vega
becomes smaller, the crude MC estimate becomes extremely noisy. As mentioned,
this is due to fact that the small value of Vega in this regime is the result of the cancellation of two
poorly correlated stochastic quantities representing the averages of the LRM estimator
over the configurations in which it has a definite (positive and negative)
sign.

As expected, due to the presence in the LRM weight of even terms in the random
increments, the efficacy of Antithetic Variables is very limited for Vega (see Table \ref{tab:vegabasket}).
In contrast, using both the
LRM weight and the Vega of the Forward as Control Variates provides sizable variance reductions both
for in the money and at the money options, with the expected remarkable efficiency gains for small
strikes. On the other hand, Importance Sampling -- although producing  smaller
variance reductions for in the money options -- becomes particularly effective for larger strikes.

In this sense, Importance Sampling and Control Variates appear to be somewhat complementary for this problem,  
and Figure \ref{fig2} displays the LRM Vega as obtained by choosing the most effective of the two techniques 
for each strike. This results in orders of magnitude savings in computer time with respect to the 
crude MC calculation.

\section*{ACKNOWLEDGMENTS}
It is a pleasure to acknowledge useful discussions with Mike Giles, Paul Glasserman,
David Shorthouse, Jacky Lee, Anton Merlushkin, Mark Stedman, and Sanjay Chawla.
The opinions and views expressed in this paper are uniquely those of the author, and do not necessarily
represent those of Credit Suisse Group.

\appendix

\section{APPENDIX: DERIVATION OF EQ.~(\ref{weight})}

First, by differentiating the joint cumulative distribution
(\ref{gaucop}) one obtains the corresponding probability density
function
\begin{eqnarray}\label{density}
P(x)&=&
\phi_N\left(\Phi^{-1}\left(M_1(x_1)\right),\ldots,\Phi^{-1}\left(M_N(x_N)\right);\Sigma\right) \nonumber \\
&\times& \,\prod_{i=1}^N \frac{m_i(x_i)}{\phi(\Phi^{-1}(M_i(x_i)))}~,
\end{eqnarray}
where $m_i(x_i) = d M(x_i)/d x_i$ is the $i$-th marginal density
function, and
$\phi_N(x_1,\ldots,x_N;\Sigma)$ is the multivariate Gaussian
density with correlation $\Sigma$.
Then, taking the logarithm of Eq.~(\ref{density}) gives
\begin{eqnarray}
\log P(x) &=& \sum_{i=1}^N \Big( \log m_i(x_i)  - \log \phi(Z_i)
\Big) - \frac{1}{2} Z(x)^T \,\Sigma \,Z(x) \nonumber
\\ &-& \frac{N}{2} \log{ 2\pi } - \frac{1}{2}\log{\left(\det
\Sigma\right)}~,
\end{eqnarray}
where we have used the explicit form of $\phi_N(Z_1,\ldots Z_N; \Sigma)$,
and the definition in Eq.~(\ref{zeta}).
Hence, the derivative with respect to $\theta$ of the latter equation, when $\partial_\theta
\Sigma \equiv 0$, can be written as in Eq.~(\ref{weight})
where $\partial_\theta Z(x)$ (\ref{dcdf}) can be obtained by deriving Eq.~(\ref{zeta}).

\bibliographystyle{wsc}
\bibliography{biblio}

\section*{AUTHOR BIOGRAPHY}

\noindent {\bf LUCA\ CAPRIOTTI} is a Vice President at Credit Suisse Group,
Investment Banking Division, where he works in the Global Modelling and Analytics Group (GMAG).
He was previously a researcher at the Kavli Institute for Theoretical Physics,
Santa Barbara, California, working in the field of High Temperature
Superconductivity and Quantum Monte Carlo methods for Condensed Matter systems.
His current interests are in the field of Computational Finance, mainly focusing
on efficient numerical techniques for Derivatives Pricing and Risk Management.

\end{document}